\documentclass{aa}  
\usepackage{array}  
\newcolumntype{C}{>{$}c<{$}}
\usepackage{adjustbox}
\usepackage[skip=0.333\baselineskip]{caption}
\usepackage{booktabs, 
             	amsmath,  
           	 siunitx}  
\usepackage{url}
\usepackage[utf8]{inputenc}
\usepackage[T1]{fontenc}

\pdfoutput=1 

\usepackage{graphicx}
\usepackage[pdftex]{}
\usepackage{xspace}

\usepackage[varg]{txfonts}

\usepackage{hyperref}

\begin{document} 
\title{Comparing the reflectivity of ungrouped carbonaceous chondrites with that of short period comets like 2P/Encke}

   \author{Safoura Tanbakouei
          \inst{1,2},
          \
           Josep M. Trigo-Rodríguez\inst{1,2},
\
J$\ddot{u}$rgen blum\inst{3},
\
Iwan Williams \inst{4},
\
 Jordi Llorca\inst{5} }

 \institute{Institute of Space Sciences (ICE-CSIC), Campus UAB, C/ Can Magrans s/n, 08193 bellaterra (Barcelona), Catalonia, Spain. \\
              \email{tanbakouei@ice.csic.es}
	         \and
Institut d’Estudis Espacials de Catalunya (IEEC), C/ Gran Capit$\grave{a}$, 2-4, Ed. Nexus, desp. 201, 08034 Barcelona, Catalonia, Spain. \\
             \email{trigo@ice.csic.es}
 \and
Institut f$\ddot{u}$r Geophysik und extraterrestrische Physik, Technische Universit$\ddot{a}$t Braunschweig, Mendelssohnstr. 3, 38106 Braunschweig, Germany. \\
              \email{j.blum@tu-bs.de}
 \and
School of Physics and Astronomy, Queen Mary, University of London, Mile End Rd. London E1 4NS, UK. \\
              \email:{i.p.williams@qmul.ac.uk}
\and
Institute of Energy Technologies, Dept. Chemical Engineering and Barcelona Research Center in Multiscale Science and Engineering, Universitat Politècnica de Catalunya- BarcelonaTech, Catalonia, Spain.\\
              \email{ jordi.llorca@upc.edu}
}

\pagestyle{plain} 
\thispagestyle{empty}
\abstract{}{The existence of asteroid complexes produced by the disruption of these comets suggests that evolved comets could also produce high-strength materials able to survive as meteorites. We chose as an example comet 2P/Encke, one of the largest object of the so-called Taurid complex. We compare the reflectance spectrum of this comet with the laboratory spectra of some Antarctic ungrouped carbonaceous chondrites to investigate whether some of these meteorites could be associated with evolved comets.}
{We compared the spectral behaviour of 2P/Encke with laboratory spectra of carbonaceous chondrites. Different specimens of the common carbonaceous chondrite groups do not match the overall features and slope of comet 2P/Encke. Trying anomalous carbonaceous chondrites, we found two meteorites, Meteorite Hills 01017 and Grosvenor Mountains 95551, which could be good proxies for the dark materials forming this short-period comet. We hypothesise that these two meteorites could be rare surviving samples, either from the Taurid complex or another compositionally similar body. In any case, it is difficult to get rid of the effects of terrestrial weathering in these Antarctic finds, and further studies are needed. Future sample return from the so-called dormant comets could be also useful to establish a ground truth on the materials forming evolved short-period comets.}{As a natural outcome, we think that identifying good proxies of 2P/Encke-forming materials might have interesting implications for future sample-return missions to evolved, potentially dormant or extinct comets. To understand the compositional nature of evolved comets is particularly relevant in the context of the future mitigation of impact hazard from these dark and dangerous projectiles.}
{}
\keywords {2P/Encke comet -- 
	meteorites --
	reflectivity --
	Taurid complex
	}
\titlerunning{Similar reflectivity of 2/PEncke comet and Ungrouped CCs}
\authorrunning{S. Tanbakouei et al. 2020}
\maketitle
Accepted for publication in A\&A on July 6, 2020

\section{Introduction}

Comet 2P/Encke, hereafter Encke, is a 4.8 km-sized active comet discovered in 1786. It has an unusual orbit with a period of 3.3 years, and is one of the largest known object in the Taurid complex (hereafter TC) \citep{10.1093/mnras/211.4.953, asher1993asteroids}. Dynamically the TC was thought to have its origin within the main asteroid belt \citep{hsieh2006population, Jewitt_2012}, but the discovery of tens of asteroids in similar orbits to the meteoroid complex suggest its origin in the disruption of a more eccentric and large comet. 
A formation scenario where 2P/Encke is one of many fragments resulting from the break-up of a much larger comet was proposed by \citet{10.1093/mnras/211.4.953} and later discussed  by \citet{asher1993asteroids} and \citet{babadzhanov2008near}. The importance of the study of the TC under an impact hazard perspective seems obvious, particularly with the recent discovery of challenging dormant comets in near-Earth space \citep{binzel2015near} that might be well exemplified by the Earth’s encounter with the challenging 2015 TB145 \citep{muller2017hayabusa, MICHELI2018265}. On the other hand, a potentially hazardous asteroid (PHA), 2008 XM1, was recently identified as one of the objects producing meteoroids dynamically associated with this complex \citep{MADIEDO2014356}. Moreover, Earth’s encounter with the TC debris has been proposed to be the source of a catastrophic aerial blast that occurred around 12 900 BC, which promoted a return to ice age conditions for $\sim$1300 years \citep{10.1111/j.1365-2966.2010.16579.x}. On the other hand, recent meteor studies have suggested that bright Taurid fireballs exhibit finite ending masses, so that there is some chance that meteorite collections contain free-delivered samples of 2P/Encke or some other members of the TC \citep{MADIEDO2014356, spurny2017discovery}.

Due to gravitational interactions with the terrestrial planets, Encke’s orbit is far from Jupiter’s control and results in a very smooth orbital evolution (see e.g. \citet{LEVISON2006161}). The properties of Encke are quite distinct compared to other objects of the Solar System \citep{boehnhardt2008photometry}. Compared to the available spectra of other cometary nuclei, 2P/Encke’s spectrum is rather typical, despite its very peculiar orbit. It is also important to remark that 2P/Encke exemplifies a most challenging hazardous body, because it has a very low albedo, reflecting only 4.6\% of the light it receives in the visible range \citep{tubiana20152p}. 

It might be surprizing that most members of the TC are asteroidal in nature, but it is probably a natural outcome of the evolution of short period comets \citep{binzel2015near, jewitt2008kuiper}. The volatile content for each one of these Taurid members was perhaps inhomogeneous, or possibly affected by other processes that extinguished their volatile content, like e.g. close approaches to the Sun or impact gardening \citep{beitz2016collisional}. Obviously, these new objects found suggest that the comet 2P/Encke could be part of a transitional and inhomogeneous progenitor, coming from a formation region in the asteroid/comet boundary \citep{hsieh2006population, hsieh2017asteroid}. In such a sense, a recent study of the reflectance spectra of the largest asteroids dynamically associated with 2P/Encke, revealed that their spectral characteristics are linked with the S taxonomic complex \citep{popescu2014spectral}. Comet 2P/ Encke has a visible spectrum, similar to other cometary nuclei and closer to D-type asteroids \citep{raponi2020infrared}, as it was also suggested from the behavior of ungrouped Tagish Lake carbonaceous chondrite \citep{hiroi2001tagish}. Such a result points towards a heterogeneous comet progenitor that after disruption produced the known PHAs on Apollo type orbits, some of them exhibiting ordinary chondrite-like mineralogy. It could have deep implications for impact hazard, but given the degree of collisional processing, we cannot discard that short-period comets suffered the implantation of foreign rocky projectiles that contribute to the modification of their surfaces and produce high-strength meteoroids when disrupted \citep{kresak1978tunguska}. We should also mention in this context that the Tunguska airblast on June 30, 1908 was initially associated with the TC \citep{trigo2017dynamic}. \citet{SEKANINA1998191} later discissed such a cometary nature with the argument that the fireball had a disruption behaviour characteristic of a stony asteroid. If the bodies forming the TC are so heterogeneous in nature, it could make room for being both scenarios correct, after all. In any case, it has been accepted that a hydrated asteroid should progressively deplete its volatiles that are mostly bound in minerals rather than forming ices \citep{trigo2015aqueous, trigo2019accretion}. In fact, most TC members behave like asteroids and do not exhibit any cometary activity at all. In this sense, we are interested in testing whether some carbonaceous chondrites (CC) could be good proxies of the reflectance properties of comet 2P/Encke. Current formation models predict that short-period comets can only stay active for a relatively short time (or number of perihelion passages), because their high internal lithostatic compression hardens the material so much that activity stops when a thin surface layer has been emitted \citep{Gundlach_2016}.  

 The reflectance spectrum of 2P/Encke typically exhibits a moderate red slope and is otherwise featureless \citep{LUU199069}. To our knowledge, it has so far not been possible to find a meteorite proxy with a similar reflectance behaviour. Consequently, the main goal of our work is to identify samples in our meteorite collections with reflectance characteristics similar to comet 2P/Encke. We consider Encke as a case study to promote studies with the aim to establish a link between short-period comets using remote sensing and laboratory studies of meteorites.

\section{Technical procedure, and sample selection}

We obtained reflectance spectra of the meteorites described in Tables 1 and 2, using the procedure described in previous work \citep{10.1093/mnras/stt1873}. Polished sections of the selected meteorites were measured at UPC, using a Shimadzu UV3600 Ultraviolet to Near-infrared (UV-Vis-NIR) spectrometer. The spectrometer diffraction- limited illumination originates from one of two lamps and passes through a variable slit, then is filtered with a grating to select the desired wavelength and afterwards is split into two alternating but identical beams with a chopper. Next, the beam interacts with the sample and is routed to the detector. The reference beam interacts with the material of the sample surface and then goes to the same detector \citep{moyano2016plausible}. Next, the beam interacts with the sample with an angle of 8$^{\circ}$ and it is later routed to the detector. The standard stage for the spectrometer is an integrating sphere with a working range in the current study of 400 to 900 nm, and operated under laboratory conditions. 
The reflectance spectrum of 2P/Encke was taken from \citep{tubiana20152p}. The spectrum was normalised to unity at 550 nm. It is important to remark that the 2P/Encke reflectance data was removed between 570 nm and 620 nm between 720 nm and 780 nm, due to several issues related with the data reduction and some artefacts in the spectra as explained in \citet{tubiana20152p}. The meteorite specimens studied in this work and compared with 2P/Encke are listed in ~Table \ref{Tab1} and ~Table \ref{tab2}. 

 \begin{table}
    \caption{List of CR, CM and CK carbonaceous chondrites that were compared to the spectrum of comet 2P/Encke in this work. The masses were taken from the Meteoritical Bulletin and denote the total mass of the meteorite.}
    \label{Tab1}
    \centering
    \begin{tabular}{C C C } 
    \toprule
    Meteorites & Group  & Mass (g) \\
    \midrule
    QUE  90355 & CM2 & 32.4 \\
    Murchison & CM2   &  \sim 100,000 \\
    EET  92159 & CR2 & 67.6 \\
   LAP  02342 & CR2 & 42.4 \\
   Renazzo & CR2 & \sim 1,000\\
  PCA  82500 & CK4-5 & 90.9 \\
  LAR  12265 thin & CK5 & 14.3\\
 ALH  82500 & CK5 & 438\\
    \bottomrule
    \end{tabular}
\end{table}

\section{Results and discussion}

In order to gain insight into the nature of evolved comets, we compared the reflectance spectra of CCs with the reflectance spectrum obtained for comet 2P/Encke \citep{tubiana20152p}. In our study, we included a significant number of NASA Antarctic CCs. Given the small terminal mass expected for meteorite-dropping bolides of the Taurid stream, we think that the Antarctic collection can provide the opportunity to identify the first meteorites associated with comets. 

\citet{gradie1986wavelength} studied the visible to NIR spectral dependence of ordinary and carbonaceous chondrites. Their models of the surface roughness indicate that the differences found between the visual phase coefficients of S- and C-class asteroids are reflecting primarily differences in composition and not necessarily differences in surface roughness. The beam incidence angle of 8$^{\circ}$ is a particular geometry that cannot be changed in our spectrometer, so we can consider that our data is obtained in almost specular geometry (see e.g. \citet{britt1988bidirectional}). These authors demonstrate that the amount of metal plays a significant role in the slope of the spectrum as we also previously discussed \citep{moyano2016plausible}. On the other hand, the incidence angle also might affect the slope of the spectrum, and the overall spectral reflectivity but has not influence in the location of absorption bands \citep{gradie1980effects}.

Our reflectance spectra show features that might be different to those obtained using remote sensing. The reason for that is that each feature depth depends on many factors, e.g., differences in observation geometry, or the grain size, shape and roughness. In addition, we recognise that the spectral slope of carbonaceous chondrites can change from red to blue depending on whether the sample is prepared as powder or as slab. The slope also changes with grains size, packing density or observation geometry as has been studied for Murchison CC \citep{cloutis2018spectral}. For these reasons, the sample reflectance slope is a multivariate problem, and it provides only a first approach to look for possible proxies of solar system primitive bodies.

Comparison of the reflectance spectrum of comet Encke with those obtained for specimens of CK, CM and CR chondrites has previously been done (see e.g. \citet{10.1093/mnras/stt1873} and \citet{tanbakouei2019mechanical}) ~(Table \ref{Tab1}). We chose groups that exhibit significant evidence for parent body aqueous alteration, or oxidation \citep{trigo2015aqueous, trigo2017dynamic}. No clear similarity with 2P/Encke’s spectrum was found as a result of such comparisons for any specimen belonging to these well-established CC groups. Then, we performed a similar one-by-one comparison with the spectra of some ungrouped CCs and found two quite reasonable matches that we describe in the next section \citep{CLOUTIS2011309} ~(Table \ref{tab2}).

 \begin{table}
    \caption{Ungrouped carbonaceous chondrites found in this work to possess a reasonable similarity to the spectrum of comet 2P/Encke. The masses were taken from the Meteoritical Bulletin and denote the total mass of the meteorite.}
    \label{tab2}
    \centering
    \begin{tabular}{C C C }
    \toprule
    Meteorites & Group  & Mass (g) \\
    \midrule
    MET  01017 & CV3-an &   238.0 \\
   GRO  95551 & C-ung   &      213.3 \\
    \bottomrule
    \end{tabular}
\end{table}

We did our best to get rid of terrestrial weathering, choosing those areas of the two meteorites that were minimally affected by the rusting effect. However, both ungrouped carbonaceous chondrites belong to the NASA Antarctic collection and probably spent a significant time in Antarctica. Consequently, we should remark the possible impact of terrestrial weathering in both spectra. There is a possibility that the origin of the spectral slope measured for the two anomalous carbonaceous chondrites is caused by some oxides formed by terrestrial weathering. We do not think this
to be the case in the areas where we collected the spectra, but if these ungrouped carbonaceous chondrites were affected by terrestrial weathering, the spectral slope could change from red to blue depending on whether the sample is prepared as powder or as slab \citep{cloutis2018spectral}.
These authors also demonstrated that the slope is very sensitive and could change with the grain size, packing density or the observation geometry. In particular, the particle size can have a strong effect on the albedo so that a body covered in fine-grained regolith with a particular phase angle has to be taken into account to do a proper comparison with the spectral properties of meteorites \citep{muinonen2010three, belskaya2010polarimetry}.

\subsection{Comparison with ungrouped carbonaceous chondrites}

The reflectance spectrum of 2P/Encke covers the full visible range from 400 nm to 900 nm on the detector. Concerning the comparison with other CC groups, the specimens belonging to most carbonaceous chondrite groups exhibited very different reflectance spectra. After many attempts among the meteorites studied, we found a likely reflectance similarity in one meteorite, named Meteorite Hills 01017 (MET 01017). This meteorite was initially characterised as a CV3 (possibly reduced) chondrite \citep{busemann2007characterization}, but due to its anomalous properties was reclassified as a CV3-an (MetBull: 88, 2001). In fact, the section used in our work shows evidence of aqueous alteration with half of its exterior having a weathered fusion crust with oxidation haloes. The section also exhibits mm-sized chondrules quite well preserved and possesses a fractured and weathered interior. It also contains metal-rich chondrules and CAIs in a dark matrix of FeO-rich phyllosilicates \citep{brearley1997disordered, krot2006timescales, trigo2015aqueous}. CV chondrites usually contain a 45\% volume of large chondrules with an average size of 1 mm, $\sim$40\% volume of dark matrix, and a large volume abundance of CAIs of $\sim$ 10\% \citep{brearley1998planetary, scott2003chondrites}. The matrices of CV3 chondrites are dominated by fine-grained olivine plus phyllosilicates and opaque materials such as metallic iron, sulphides and magnetite \citep{burbine2001k}. Metal grains typically range in size from 50 to 500 nm. Sulphides, up to $\sim$200 $\mu$m in size, are usually associated with metal, and act as opaque material as well.

Another sample with a spectral behaviour similar to 2P/Encke is Grosvenor Mountains 95551 (GRO 95551). This meteorite is an ungrouped CC, an unusual metal-rich breccia formed by two main types of clasts, chondritic and achondritic. The chondritic clasts consist of a variety of chondrules and chondrule fragments. The mean olivine composition of GRO 95551 is Fa$_{1.3}$ (21 analyses) with a range of Fa$_{0.7-3.5}$ and pyroxene ranges Fs$_{0.7-39.6}$, Wo$_{0.7-1.6}$ \citep{WEISBERG2015269}. The meteorite is anomalous with a high metal content and some petrographic peculiarities, like e.g. the presence of large silicate nodules, resembling Bencubbin \citep{KALLEMEYN1978507, weisberg2001new}. The specimen is of interest, because it might be representative of the materials stemming from processed objects \citep{trigo2015aqueous, brearley1998planetary, moyano2015comparative}.

\begin{figure}
   \centering
\includegraphics[width=9cm]{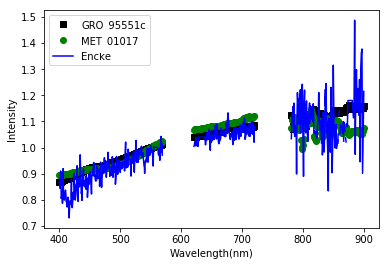}
      \caption{Reflectance spectra from 400 to 900 nm of the two ungrouped \textit{carbonaceous} chondrites compared with the spectrum of comet 2P/Encke (blue line) (from \citep{tubiana20152p}). MET 01017 belongs to the CV3-an chondrite group and GRO 95551 is an ungrouped carbonaceous chondrite. All spectra were scaled and normalised to 1 at 550 nm.
              }
          \label{fig1}
   \end{figure}

As we can see in Fig. \ref{fig1}, at first sight the spectral slope of the two studied meteorites is compatible with that of 2P/Encke. The spectrum of the comet shows shallower slope above 800 nm. However, as we explain in Section 2, our reflectance spectra were taken under a beam angle of 8$^{\circ}$ so that we need to recognise that it implies a significant limitation in the comparison with remote sensing data, particularly in reference to slope. Said that, the spectra of the comet is similar to the spectra of primitive asteroids and seems to be weird not having some evidence, even when it could be scarce in meteorite collections. It is similar to the spectra of primitive asteroids, which are believed to be associated with CCs \citep{tubiana20152p}. ~Figure \ref{fig1} shows that both ungrouped CC samples and comet Encke exhibit a common absorption feature at 500 nm ~(Table \ref{tab3}). From the spectral similarities, a tentative link between 2P/Encke, MET 01017 and GRO 95551 can be made. It is noticeable to know that ungrouped carbonaceous chondrites have highly variable spectral properties and that the surface of comet Encke was subjected to significant changes due to solar heating \citep{tubiana20152p}.

 \begin{table}
    \caption{The location of the main absorption bands in the two ungrouped CC meteorites and their mineral assignments \citep{cloutis2012spectral, cloutis2012spectralcv, trigo2014uv, weisberg2015petrology}.}
    \label{tab3}
    \centering
    \begin{tabular}{C C C } 
    \toprule
    Meteorites & Absorption  band (nm) & Mineral \\
    \midrule  
MET  01017  &      495    &          Olivine     \\
                         &       665 &      Magnetite/ maghemite  \\ 
                         &       700-720   &    Serpentine (phyllosilicates) \\
		    &       850           &        Olivine \\
		    &       875         &          Orthopyroxene  \\
\midrule
GRO  95551    &         470   &        Olivine             \\
                          &     665    &     Mgnetite/ maghemite      \\
 		    &         710  &    Serpentine (phyllosilicates) \\
		     &         850  &    Olivine \\
    \bottomrule
    \end{tabular}
\end{table}

\citet{WEISBERG2015269} showed the similarities of Bencubbin and GRO 95551, but they are different in olivine compositions, oxygen isotope compositions and presence of interstitial sulfides, and also in the siderophile element compositions of the metal \citep{WEISBERG2015269}. The siderophile element composition of GRO 95551 is closer to that of ALH 85085 than to that of Bencubbin \citep{kallemeyn2000composition}. There are many of the smaller non-spherical objects in GRO 95551 that show the texture of chondrules. The lack of matrix in GRO 95551 could be a sign of rapid accretion, with significant thermal processing \citep{metzler2012ultrarapid}. Obviously some of these features could be also the consequence of shock processing and/or heating in close approaches to the Sun \citep{delbo2014thermal}.

\begin{figure}
   \centering
  \includegraphics[width=9cm]{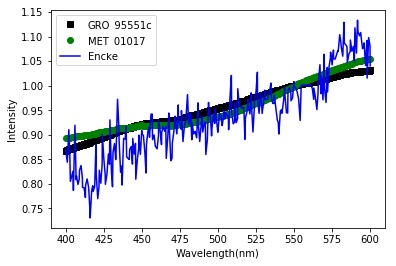}
      \caption{Reflectance spectra from 400 to 600 nm of the two ungrouped \textit{carbonaceous} chondrites compared with the spectrum of comet 2P/Encke (from \citep{tubiana20152p}). }
          \label{fig2}
   \end{figure}

Concerning the reflectance spectra of MET 01017, a weaker absorption feature around 470-520 nm can also be noticed and appears to be compatible with a weak feature also found in the 2P/Encke reflectance spectrum ~(Fig. \ref{fig2}). It is important to consider that comet surfaces are probably extensively processed due to solar heating and collisional gardening \citep{trigo2015aqueous}. In addition to the reflectance data, additional evidence has been recently obtained by fireball networks, suggesting that Taurid bolides associated with the largest 2P/Encke meteoroids could produce meteorites and even might contribute significantly to the impact hazard \citep{MADIEDO2014356, 10.1111/j.1365-2966.2010.16579.x}.

For the characteristics of the TC, we expect a quite heterogeneous progenitor, and as a consequence of its short orbital period as an evolved comet, we could also propose a brecciated nature due to collisional gardening \citep{trigo2015aqueous}. Therefore, a big diversity of anomalous meteorites could be expected from a disrupted and brecciated object. Just as an example of the outcome of a body subjected to strong collisional processing, we could cite the heterogeneity found in Almahatta Sitta ureilite body, containing materials from many different rocky projectiles \citep{bischoff2010asteroid}.

\subsection{Comparison with grouped carbonaceous chondrites}

\begin{figure}
   \centering
  \includegraphics[width=9cm]{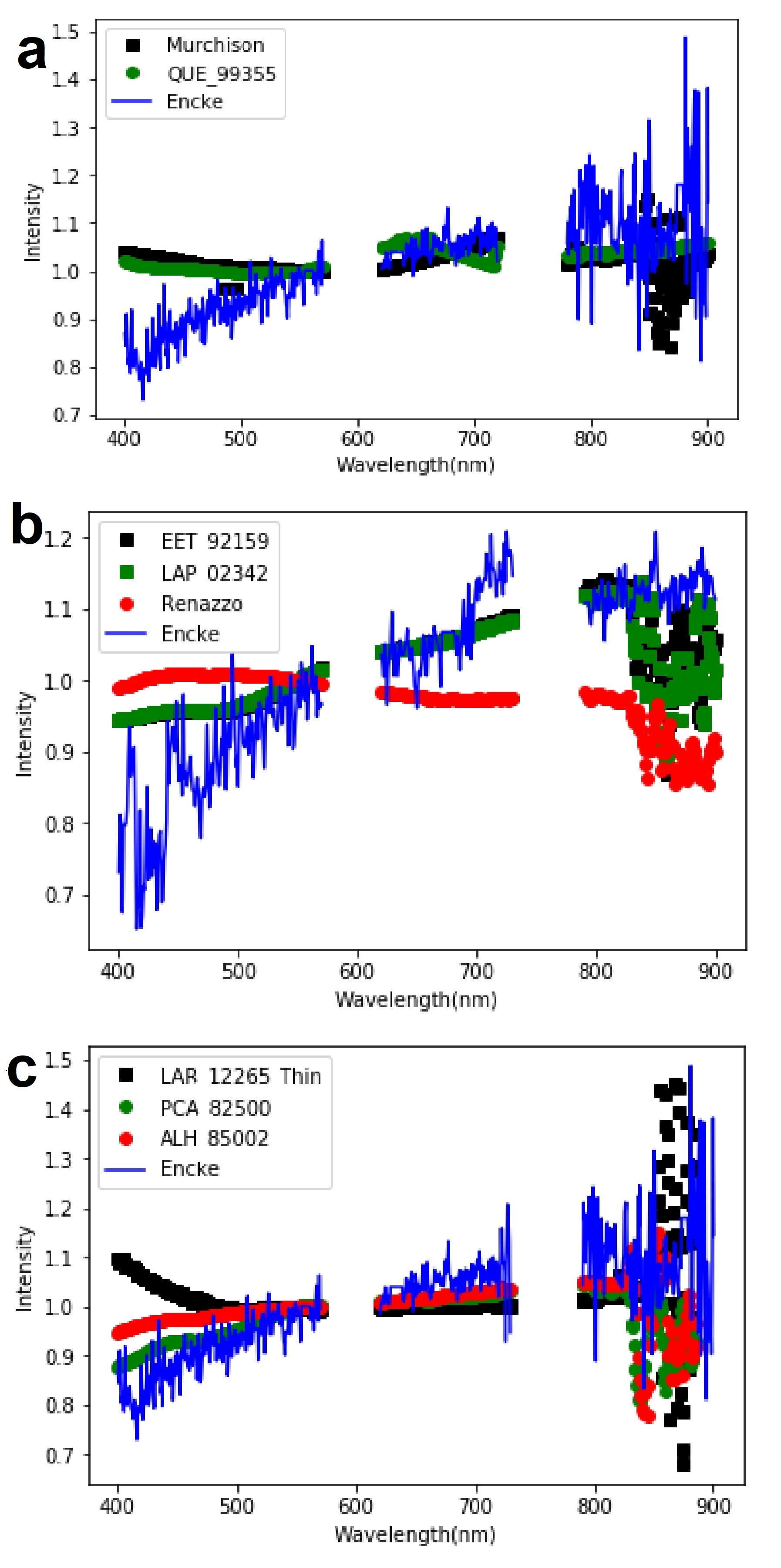}
      \caption{Comparison of reflectance properties of comet 2P/Encke with selected specimens of the CC groups. Two CM2 chondrites are compared in (a), three CR2 chondrites in (b) and three CKs in (c). All data were normalized to 1 at 550 nm.  }
          \label{fig3}
   \end{figure}

Tubiana et al. found that the 2P/Encke reflectance spectrum is similar to the characteristics of primitive asteroids \citep{tubiana20152p}, which are supposed to be correlated with CCs \citep{trigo2015aqueous}. For that reason, we decided to establish a comparison with the rest of the CC spectra belonging to the chondrite groups of CK, CR and CM. All the spectra were obtained in the laboratory covered the 400-900 nm wavelength range and are included in ~Fig.\ref{fig3}.

CM chondrites probably originated from a hydrated C-rich asteroid \citep{trigo2019accretion, RUBIN20072361}. Aqueous alteration of metal and silicate phases in that volatile-rich body have formed phyllosilicates and iron alteration minerals \citep{rivkin2002hydrated}. Absorption band features around 700 nm in Fig. 3a are associated with hydrated minerals, so they could disappear after heating the material by, for example, collisional processing \citep{hiroi1996thermal}. As \citet{tubiana20152p} mentioned, by looking at meteorite samples, we think that the surface spectra of comets and primitive asteroids could be not representative of their interiors. Despite the low overall reflectance, the other absorption features in the 380 to 450 nm region can indicate the presence of insoluble organic matter, or a metal–O charge transfer absorption \citep{cloutis2008ultraviolet}, while the absorption features in the 450 to 480 nm range exhibit the presence of oxidised Fe species like hematite and goethite that produce an absorption band more extended than that of magnetite. It is also evident the presence of an extended olivine band centered in 495 nm that is also visible in MET 01017. The CM carbonaceous chondrites we studied here, Murchison and QUE 99355, are CM2 chondrites. According to the taxonomic classification scheme \citep{tholen1984asteroid}, there is a relation between the CM2 chondrites and asteroids, corresponding to an absorption band in the 700-750 nm region and an absorption feature near 900 nm (another absorption feature is in the 1100 nm region, although the spectral coverage of this wavelength region is not available). Aqueous alteration of metal and silicate phases in that volatile-rich body have formed phyllosilicates and iron alteration minerals \citep{zolensky1988aqueous, brearley1998planetary, krot2006timescales, alexander2012provenances}.

The CR chondrites belong to one of the major CC groups, with abundant free metal and magnetite that are a consequence of aqueous alteration \citep{kallemeyn1994compositional, CLOUTIS2012586}. Olivine is the dominant silicate in the CR2 and CR3 chondrites \citep{brearley1998planetary}. The spectral reflectance of the CR2 chondrite Renazzo shows that aqueous processes produce the alteration of olivine and pyroxene to phyllosilicates \citep{tholen1984asteroid}. Concerning the asteroid parent body of the CR chondrites, previous authors have not found any good match among the C, G, B, and F asteroids \citep{CLOUTIS2012586, hiroi1993evidence, hiroi1997characterization}. It is expected that the typical spectral features of the CR2 parent bodies are Fe$^{2+}$ phyllosilicate, low-Fe olivine absorption features, a moderate overall reflectance ($\sim$8–12\%), the presence of weak silicate absorption bands around 900 and 1100 nm, a weak olivine absorption band at 900 nm, and absent or quite weak Fe$^{3+}$- Fe$^{2+}$ phyllosilicate absorption bands in the 650-750 nm wavelength range \citep{CLOUTIS2012586}. The CR2 carbonaceous chondrites selected here to compare with the spectral properties of comet Encke are EET 92159, LAP 02342 and Renazzo. ~Figure. \ref{fig3}b clearly shows a very different spectral behaviour for the CR chondrites compared with 2P/Encke.

Concerning the spectral reflectance data for the CK group, we chose three specimens: LAR 12265 CK5, PCA 82500 CK4/5 and ALH 85002 CK5. This group is of interest here, because a previous comparison between CCs and the Eos asteroid family members provided a good match \citep{mothe2008mineralogical}. ~Figure. \ref{fig3}c exemplifies that there is probably not a compositional similitude between comet Encke and the CK group. All, the spectra of the CK meteorites do not show a shallow slope beyond 800 nm in contrast to Encke’s spectrum, which is relatively flat beyond the 800 nm region. 

To explain in which way such extremely different objects as meteorites and comets may belong to the same TC is of utmost importance. It is well known that meteorites possess typical cohesive (tensile) strengths of several MPa (see, e.g.,  \citet{blum2006physics} for a compilation), whereas comets are extremely weak bodies. The latest estimate of the tensile strength of comets is only $\sim$1 Pa \citep{skorov2012dust, blum2014comets, attree2018tensile, blum2017evidence} for length scales above $\sim$1 cm and $\sim$1 kPa for length scales below that \citep{moyano2016plausible}. The latter can be measured in meteor streams \citep{trigo2006strength, trigo2007erratum}), whereas the former is a requirement for the comet to become active \citep{gundlach2020activity}. Besides this vast difference in cohesion, meteorites and comets also differ in porosity. While meteorites exhibit very low porosities of $\leqq$20\% \citep{britt2003stony}, comets must be highly porous, with porosities of $\sim$60-80\% \citep{blum2006physics, kofman2015properties, patzold2016homogeneous, fulle2016comet, herique2019homogeneity}. 

All primitive bodies in the Solar System date back to the formation era in which protoplanetary dust evolved into planetesimals. In the past years, evidence was gathered that comets in general \citep{skorov2012dust, blum2014comets}, comet 67P/Churyumov-Gerasimenko \citep{blum2017evidence, fulle2016comet} in particular and Kuiper-Belt object (486958) Arrokoth \citep{mckinnon2020solar} were formed through the gentle gravitational collapse of a cloud of cm-sized “pebbles” (dust agglomerates). If we also apply such a formation mechanism for the progenitor of the members of the TC, we can state that its maximum size is constrained by several processes: (i) Collisional destruction of the pebbles during the gravitational collapse. Experimental and numerical work has shown that a maximum radius of the final object of $\sim$50 km is possible, before the pebbles would become collisionally fragmented \citep{jansson2017role}, which would, thus, lead to a cohesive strength so high that dust activity would be impossible. (ii) Lithostatic destruction or deformation of the pebbles inside the planetesimal. Given the cohesive strength of the pebbles of $\sim$ 1-10 kPa \citep{blum2006physics} and the volume-average lithostatic stress inside a body of radius R of 
\begin{equation}
\frac{4}{15} \pi \rho^2 G R^2
\end{equation}
with $\rho \approxeq$ 500 kg $m ^{-3}$ \citep{patzold2016homogeneous} and G being the mass density of the body and the gravitational constant, we get stresses of 
\begin{equation}
4 (\frac{R}{1 km})^2   Pa
\end{equation}
Thus, the maximum radius before the internal stresses destroy the pebbles is $R \approxeq$16-50 km. (iii) However, even before the lithostatic stress can destroy the pebbles, the tensile strength may rise to values that would render activity impossible. It was shown experimentally that about 2.5-3\% of the compressive stress remains in the body as tensile strength, even after the source for the stress has disappeared \citep{blum2014comets}. Setting an upper limit of 1 Pa for the tensile strength of an active comet, we get an upper limit for the progenitor body of $R \approxeq $3 km . We conclude here that the progenitor body of the Taurid complex and, thus, comet 2P/Encke and the source of the meteorites, was a rather small body. As it is the fate for small bodies in the asteroid belt to eventually get collisionally disrupted and reaggregated \citep{michel2001collisions, beitz2016collisional, schwartz2018catastrophic}, the progenitor planetesimal must have originated in a heavily underpopulated region of the young Solar System. 

Although the modelling of the dynamical evolution from planetesimal to comet and meteorites is beyond the scope of this paper, we will try to sketch one possible scenario: during the evolutionary process of the Solar System, a high-velocity collision of the planetesimal with a much smaller impactor must have locally compacted its near-surface regions \citep{beitz2016collisional, beitz2013experiments}. To achieve porosities $\leqq 20\% $ (measured for meteorites \citep{britt2003stony}), impact speeds of typically 1 km$ s ^{-1}$ or above are required. If the impactor was sufficiently small, the collision was sub-catastrophic and the comet nucleus remained the largest intact post-collisional residue that suffered neither substantial compaction nor heating \citep{beitz2013experiments}, whereas sufficient compaction occurred at the impact site to create the low-porosity meteoritic matter \citep{beitz2016collisional}. Such evolutionary scenario could be consistent with a significant number of ungrouped CCs available in meteorite collections (about a 3\% of all CCs according to the Meteoritical Bulletin Database: https://www.lpi.usra.edu/meteor/metbull.php). These ungrouped CCs are difficult to assign to CC groups because they have experienced significant thermal and aqueous alteration producing volatile element depletions and altering the organics forming their matrices \citep{alexander2007origin, trigo2015aqueous}, as we could expect to result from the heat and compaction produced by impact shock that in addition could explain their peculiar bulk elemental chemistry, with depletion of moderately volatile elements. In fact, the meteorite matrices exhibit the typical 1300 cm$^{-1}$ disorder band assigned to graphite (see eg. \citep{larsen2006micro}). Whether such a scenario is plausible, or even likely, remains the task of future studies.

\section{Conclusions}

In our search for good meteorite proxies of short period comets, we compared the reflectance spectrum of comet 2P/Encke with the spectral characteristics of several CCs. The absence of a clear spectral match between 2P/Encke and the previously cited CC groups might be not so surprising, given the diversity of rock-forming materials that we expect form C-rich bodies in the Solar System. In that sense, the new results achieved by the OSIRIS-REx mission to asteroid (101955) Bennu clearly demonstrate that some asteroids are active sources of particles, probably evolving not very differently to the apparently inactive members of the TC \citep{lauretta2019episodes}. These so-called “transitional asteroids” produce meteoroid ejections and could possess evolved surfaces exhibiting further processing influenced by localized ice sublimation, phyllosilicate dehydration, thermal stress fracturing, and secondary impacts by their own ejected debris \citep{lauretta2019episodes}. Searching for ungrouped CCs that have experienced thermal processing makes sense as the reflectance spectra of evolved comets, like 2P/Encke, is modelled by the presence of thermally modified minerals, like e.g. amorphous carbon, and mixtures of amorphous and crystalline silicates \citep{kelley2006spitzer}. We found that the most common and well established aqueously altered CC groups do not match the spectrum of comet 2P/Encke, but two anomalous meteorites might be good candidates to be associated with evolved comets. Consequently, we have reached the following conclusions:

   \begin{enumerate}
      \item  Despite of being almost featureless spectra, some spectral similitudes between
our two ungrouped CC and the spectrum of 2P/Encke were found. Olivine, phyllosilicates
and Fe oxides bands are present that, together with a similar spectral slope, are
reasonably matching the VIS-NIR reflectance spectrum characteristics of this short-period
comet. In any case, we recognise that additional work is needed to establish a more precise link. Fresh meteorites or sample-returned materials unaffected by terrestrial weathering processes could be particularly useful in that regard.
 
     \item Such a spectral match is not obvious for other CC groups. Comparison with CM, CR and CK chondrites exhibits no evidence of compositional similarities with comet 2P/Encke. 

    \item   Further search of the reflectance spectra of other CCs could provide additional insight into the real nature, composition and physical properties of periodic comets. Given the low albedo and overall dark nature of the surfaces of these objects, the characterisation of their rock-forming materials, like is being currently achieved by sample-return missions, is of seminal importance to mitigate future Earth encounters with these dangerous projectiles.

\item Finally, we think that a space mission to study the complex of bodies dynamically associated with comet 2P/Encke or other short period comets is of key relevance. That mission should include sample return, to recognise and understand the materials forming not only this object, but also those, probably active asteroids associated with the TC.
	\end{enumerate}

\begin{acknowledgements}
  JMTR, and ST acknowledge financial support from the Spanish Ministry (PGC2018-097374-B-I00, PI: JMTR). JLL is grateful to ICREA Academia program and funding from Generalitat de Catalunya (2017 SGR 128). US Antarctic meteorite samples are recovered by the Antarctic Search for Meteorites (ANSMET) program which has been funded by NSF and NASA, and characterized and curated by the Department of Mineral Sciences of the Smithsonian Institution and Astromaterials Acquisition and Curation Office at NASA Johnson Space Center. We thank these institutions for providing the Antarctic meteorites studied here. JB thanks the Deutsches Zentrum ff$\ddot{u}$r Luft- und Raumfahrt (DLR) and the Deutsche Forschungsgemeinschaft (DFG) for their continuous support.

\end{acknowledgements}

\bibliographystyle{aa}
\bibliography{references}

\listofobjects

\end{document}